\documentclass[aps,preprint]{revtex4}%
\usepackage{amsfonts}
\usepackage{amsmath}
\usepackage{amssymb}
\usepackage{graphicx}%
\begin{document}
\title{Weibel instability in relativistic asymmetric electron positron \ plasma}
\author{}
\author{Zahida Ehsan${}^{1\ast}$ and N. L. Tsintsadze${}^{2}$ }
\affiliation{$^{1}$COMSATS Institute of Information Technology, Lahore 54000, Pakistan.}
\affiliation{$^{2}${Institute of Physics, Tbilisi, Georgia.}\footnote{For correspondence:
ehsan.zahida@gmail.com}}
\date{\today}

\begin{abstract}

\end{abstract}
\begin{abstract}
We consider a situation in when the interaction of relativistically intense EM
waves with an isotropic electron positron plasma takes place, i.e.,we consider
short pulse lasers with intensity up to 10$^{21}$ W/cm2, in which the photon
density is of the order of 10$^{30}$cm$^{3}$ and the strength of electri field
E = 10$^{9}$ statvolt/cm. Such a situation is possible in astrophysical and
laboratory plasmas which are subject to intense laser radiation, thusleading
to nonthermal equilibrium eld radiations. Such interaction ofthe superstrong
laser radiation with an isotropic pair plasma leads to the generation of low
frequency electromagnetic EM waves and in particular a quasistationary
magnetic eld. When the relativistic circularly polarized transverse EM wave
propagates along z-axis, it creates a ponderomotive force, which a ects the
motion of particles along the direction of its propagation. On the other hand,
motion of the particles across the direction of propagation is de ned by the
ponderomotive potential. Moreover dispersion relation for the transverse EM
wave using a special distribution function, which has an anisotropic form, is
derived and is subsequently investigated for a number of special cases. In
general, it is shown that the growth rate of the EM wave strongly depends upon
its intensity.

\end{abstract}
\maketitle

\section{Introduction }

As proposed by Stephan Hawking in his path breaking paper [1], that the
objects like black hole would emit radiations, now famously known as Hawking
radiations.These radiations are very high energetic photons, which can be
converted into electron positron pair for a short time though via the
well-known pair production. In which gamma rays photons are converted into the
electron and it's anti particle positron via Einsteins mass energy equation
$E=Mc^{2}$. However it is still unclear how the pair is formed in
astrophysical environment since there are no atomic nuclei as is believed to
be the necessary for the recoil in the pair production. At adequate
temperature near $10^{9}$ K, electron and positrons appear spontaneously i.
e.,
\[
\gamma+\gamma\rightleftharpoons e^{+}+e^{-}%
\]

A pulsar is an object that emits extremely regular pulses of radio waves. It
is generally believed that a pulsar is a rapidly rotating Neutron Star which
is extremely hot along with intense magnetic field. Because of the pulsar
co-rotation with its magnetic field an electric field is generated, which has
non zero component along magnetic field. The electric field ejects particles
from pulsar surface and accelerates them up to relativistic velocities. The
particle moving along curved magnetic field lines radiates $\gamma$ quanta and
when its kinetic energy, $\varepsilon_{\gamma,}$ exceeds twice the electron
rest energy $2m_{0}c^{2}\left(  \varepsilon_{\gamma}>2m_{0}c^{2}\right)  $,
$\gamma$ quanta decays into an electron -positron pair. This pair is also
accelerated in the electric field and $\gamma$ quanta appears again, which
decays in to an electron-positron pair. This fills the magnetosphere of the
pulsar with the relativistic electron-positron plasma, which in turn screens
out the electric field generated by the pulsar rotation. 

Electron-positron plasmas are believed to be an important ingredient of the
early universe and in astrophysical objects such as pulsars, supernova
remnants, active galactic nuclei and in gamma-ray bursts. In such extreme
environments, the electron-positron pairs may be created by collisions between
particles that are accelerated by electromagnetic and electrostatic waves
and/or by gravitational forces. In pulsar environments, there is also
possibility of pair creations via high energy curvature radiation photons that
are triggered by charged particles streaming along the curved magnetic field,
with a resulting collection of positrons at the polar cups of the pulsar. High
energy laser plasma interactions and fusion devices on Earth also constitute a
source of electron-positron plasmas. experiments with petawatt lasers (with
intensities exceeding $10^{20}$ $W/cm^{2})$ have demonstrated the production
of MeV electrons and evidence of positron production via electron collisions.
Positrons are also believed to be created in post-disruption plasma in large
tokamaks through collisions between MeV electrons and thermal particles.

Recently, there has been a great deal of interest in the study of
electron-positron plasma. In fact, in the entire astrophysics the matter is
usually at a very high temperature. When the plasma temperature becomes of the
order of (or larger than) the rest energy of electrons $m_{0}c^{2}$, it
becomes relativistic. In this relativistic regime the processes of creation
and annihilation of electron-positron pairs become important. The equilibrium
is established on account of the physical mechanism quite similar to that of
the Rayleigh Jeans law blackbody radiation. Furthermore, at thermodynamical
equilibrium the chemical potential of and electron-positron gas is identically
zero. As we have already mentioned, at sufficiently high temperatures, say
from $10^{9}$ K,the electrons and positrons appear spontaneously in any region
of the space. They are created by the temperature radiation itself, i.e.

There are two types of relativistic regimes in plasma. When a plasma is
subjected to a strong electromagnetic field, the plasma particles may acquire
relativistic velocities or the thermal energy of the plasma particles become
of the order of or larger than the rest mass energy. Objects like pulsars,
quasars, active galactic nuclei and black holes are characterized by such
relativistic effects. In laboratories, such plasmas may be created either by
heating a gas to very high temperatures or by the impact of a high-energy
particle beam or by high-intensity ultrashort laser pulses. Plasmas hot enough
for electrons to be relativistic exist in different astrophysical environments
but for particles heavier than electrons such environments are rare since more
energy is required to accelerate them to a significant fraction of the speed
of light. For relativistic plasmas the thermal energy of particles is greater
than the rest mass energy i.e., $3/2K_{B}T>m_{0}c^{2}$. So a relativistic
plasma with thermal distribution has temperature greater than $511keV$.

Weibel instabilities in electron-ion plasmas have been widely studied for both
unmagnetized and magnetized plasmas.Such instabilities are usually referred to
as whistler instabilities in magnetized electron-ion plasmas, when the wave
vectors are parallel to the background magnetic field $B_{0}$, and the
temperature perpendicular to $Be$ is higher than that parallel to $B_{0}$. The
Weibel instabilities in relativistic electron-ion plasmas have also been
investigated by many authors. While the analyses for weakly relativistic and
unmagnetized plasmas are not relevant for our purposes, detailed properties of
Weibel- like instabilities in fully relativistic magnetized plasmas have been
investigated, to our knowledge, only for a particular choice of distribution
function,38 which allows no spread in perpendicular momentum. Furthermore, due
to mass symmetry, the stability properties for the electron- positron plasmas
can be significantly different from those for the electron-ion plasmas. For
example, when the electrons and positrons have identical distribution
function, the dispersion relation is the same for the right-hand and left-hand
circularly polarized modes; therefore, the waves and instabilities can have
arbitrary polarizations. In this paper, we study the linear stability
properties of the Weibel instability in relativistic magnetized
electron-positron-pair plasmas,with wave vector parallel to $B_{0}$. The
instability in the ultrarelativistic regime, with the typical Lorentz factor

The interaction of a plasma with short high intensity laser pulses can lead to
the magnetic part of the Lorentz force on the electrons to become as important
as the electric part leading in turn to a self-generated magnetic field.
Subsequently, it is expected that this self-generated magnetic field will
significantly change the pattern of the nonlinear laser pulse interaction with
the plasma. The relativistic non- linearities introduced by the magnetic field
interaction are of general interest in relation to the field of an ultrastrong
EM wave propagating in plasmas and in pulsar like media.

Recent developments in astronomical and astrophysical observations have
revealed that our universe is full of enigmatic phenomena, such as jets,
bursts, flares, etc. It is possible now to study and simulate extremely
complex astrophysical phenomena supernova explosion, etc. in laboratories
using intense and ultraintense lasers. Intense lasers have been used to
investigate hydrodynamics radiation flows, opacities, etc., related to
supernova explosions, giant planets, and other astrophysical systems. Thus the
study of the properties of such radiation strong laser pulse, nonthermal
equilibrium, cosmic field radiation, etc. in plasmas is of vital importance.
In the field of superstrong femtosecond pulses having power of $10^{23}%
-10^{24}$ $W/cm^{2}$, it is expected that the character of the nonlinear
response of of the medium will radically change due to relativistic effects.

Recently the properties of a relativistically intense EM wave in
electron-positron plasmas were investigated. Interest in such investigations
arises because after the Big Bang, matter constituted of electrons, positrons,
and photons almost in thermal equilibrium with one another at a temperature
much higher than $mc^{2}$. On the other hand, we now know that
electron-positron plasmas constitute pulsar atmospheres, accretion disks,
active galactic nuclei, and black holes.

It was shown that in the case of relativistically intense circularly polarized
EM waves propagating into a plasma, the perpendicular momentum $p=e_{\alpha
}A/c$ $A$ is the perpendicular component of the vector potential of the pump
EM waves relative to the propagation direction; $\alpha$ stands for the
particle species can be much larger than perpendicular component of the
thermal momentum of particles. However the momentum of particles along the
direction of propagation of EM waves remains thermal. This effect modifies the
distribution function which must be relativistic, and should take into account
the fact that in the perpendicular direction the particle momentum due to the
electromagnetic field be dominant over the thermal momentum in this direction.
Such a distribution is valid for arbitrary parallel temperature and takes into
account the anisotropy in the parallel and perpendicular direction.

The manuscript is organized as follows: basic equations and the dispersion
relations for the weible instability in electron-positron ions are presented
in section II. quantitative analysis and conclusions are given in Sec. III.

\section{Dispersion relation and distribution function}

We present a new physical concept for the case when the interaction of
relativistically intense EM waves with an isotropic plasma take place, i.e.,
we consider short pulse lasers with intensity up to $10^{21}$ $W/cm^{2}$, in
which the photon density is of the order of $10^{30}cm^{-3}$ and the strength
of electric field $E$ $10^{9}$ statvolt/cm. Such a situation is possible in
astrophysical plasmas and in laboratory plasmas which are subject to intense
laser radiation, thus leading to nonthermal equilibrium field radiations. Here
we consider the generation of low frequencies of the transverse EM wave by
taking stationary ions $\left(  m_{i}\rightarrow\infty\right)  $ and take only
a single active component of relativistic electrons. However, the relativistic
EM wave interaction with the plasma leads to the plasma becoming anisotropic
in the manner described in the preceding. The linear dispersion relation for
circularly polarized electromagnetic wave in an unmagnetized relativistic
plasma propagating along the z-axis is given by%
\begin{equation}
0=1-\frac{c^{2}k_{z}^{2}}{\omega^{2}}-\sum_{j}\frac{\omega_{p_{j}}^{2}}%
{\omega^{2}\widehat{\gamma}_{\perp_{j}}}\int\frac{d^{3}p}{\gamma_{j}}\left(
\frac{p_{\perp(j)}}{2}\right)  \times\left[  \frac{\partial}{\partial
p_{\perp(j)}}+\frac{k_{z}p_{\perp(j)}}{m}\frac{1}{\left(  \gamma\omega
-\frac{k_{z}p_{\perp(j)}}{m}\right)  }\frac{\partial}{\partial p_{\parallel
(j)}}\right]  F\left(  p_{\perp(j)}^{2},p_{\parallel(j)}\right)  \tag{1}%
\end{equation}
where $j$ represents electron and positron species. Nonrelativistic plasma
frequncy is defined as $\omega_{pj}=\left(  \frac{4\pi n_{0j}e^{2}}{m_{j}%
}\right)  ^{1/2}$, and relativistic mass factor is given as $\gamma
_{j}=\left[  1+\left(  p_{\perp(j)}^{2}/m_{j}^{2}\right)  +\left(
p_{\parallel(j)}^{2}/m_{j}^{2}\right)  \right]  ^{1/2}.$ In (1) $F(p_{\perp
(j)}^{2},p_{\parallel(j)})$ is the normalized distribution function of the
relativistic $j$ species.

Here we consider propagation of a superstrong circularly polarized EM wave
along z-axis and $\overrightarrow{E}(E_{x},E_{y},0)$. In such scanrio, the
lighter charged particles electrons and positrons obtain an additional
perpendicular (relative to the propagation of waves) momentum which is defined
by the amplitude of EM waves and is given as $\overrightarrow{\widetilde{p}%
}_{\perp(j)}=\left(  e_{j}/c\right)  \overrightarrow{A}_{_{\perp}}$, where
$\overrightarrow{A}_{_{\perp}}$ is the vector potential. The total
perpendicular momentum can be expressed as%
\begin{equation}
\overrightarrow{\widetilde{p}}_{_{\perp}(j)}=\overrightarrow{p}_{_{\perp}%
(j)}^{th}+\left(  e/c\right)  \overrightarrow{A}\tag{2}%
\end{equation}
here $\overrightarrow{\widetilde{p}}_{_{\perp(j)}}$ and $\overrightarrow
{p}_{_{\perp}(j)}^{th}$ denote the field and perpendicular thermal momentum of
the $j$ particles, respectively, which for nonrelativistic temperature are
given as%
\begin{equation}
\left\langle (\overrightarrow{\widetilde{p}}_{\perp(j)})^{2}\right\rangle \sim
m_{j}^{2}U_{pond}\tag{3}%
\end{equation}%
\begin{equation}
\left\langle \left(  \overrightarrow{p}_{_{\perp(j)}}^{th}\right)
^{2}\right\rangle \sim2m_{j}T_{_{\perp(j)}}\tag{4}%
\end{equation}
$U_{pond}=\left(  e_{j}A_{\perp}/m_{j}c^{2}\right)  ^{2}$ is the ponderomotive
potential. Under the condition $U_{pond}\gg(2T_{_{\perp(j)}}/m_{j}c^{2}),$ the
perpendicular thermal momentum of electrons and positrons can be ignored. So
thermal distribution function of the $j$ species is given by $p_{_{\parallel
(j)}}$ only and we have $(\overrightarrow{\widetilde{p}}_{\perp(j)})=\left(
e_{j}/c\right)  \overrightarrow{A}_{\perp}$. For such a case the full
equilibrium distribution function can be given as%
\begin{equation}
F\left(  p_{\perp(j)}^{2},p_{_{\parallel(j})}\right)  =\frac{1}{2m_{j}%
cK_{1}\left(  \beta_{j}\right)  }\delta\left(  \vec{p}_{\perp(j)}-\frac{e_{j}%
}{c}\vec{A}_{\perp}\right)  \exp\left(  \frac{-c\left(  m_{j}^{2}%
c^{2}+p_{_{\perp}(j)}^{2}+p_{_{\parallel(j)}}^{2}\right)  ^{\frac{1}{2}}%
}{T_{z(j)}}\right)  \tag{5}%
\end{equation}
where $K_{1}$ is the MacDonald function and $\beta_{j}=m_{j}c^{2}/T_{z(j)}$.
Substitution of Eq. (5) into (1), and integration over $p_{\perp}$ gives us
\begin{equation}
0=1-\frac{c^{2}k_{z}^{2}}{\omega^{2}}-\sum_{j}\frac{\omega_{p_{j}}^{2}}%
{\omega^{2}}\int_{-\infty}^{+\infty}\frac{d^{3}p}{\gamma_{0(j)}}%
dp_{\parallel(j)}\times\left[  1-\frac{p_{_{\perp0(j)}}^{2}\left(  \omega
^{2}-c^{2}k^{2}\right)  }{2m_{j}^{2}c^{2}\left(  \gamma_{0(j)}\omega
-\frac{k_{z}p_{\perp(j)}}{m_{j}}\right)  ^{2}}\right]  ,\tag{6}%
\end{equation}
where $p_{\perp0(j)}^{2}=\ \frac{e_{j}}{c}\vec{A}_{\perp}$ and $\gamma
_{j0}=\left[  1+\left(  p_{\perp0(j)}^{2}/m_{j}^{2}\right)  +\left(
p_{\parallel(j)}^{2}/m_{j}^{2}\right)  \right]  ^{1/2}.$ In Eq.(6) we have
used $\partial\gamma_{j}/\partial p_{_{\parallel,\perp(j)}}=p_{_{\parallel
,\perp(j)}}/\gamma_{j}m_{j}^{2}c^{2}.$ The parallel distribution function
$F(p_{\parallel})$ is given as%
\begin{equation}
F(p_{_{\parallel(j)}})=\frac{1}{2m_{j}cK_{1}\left(  \beta_{j}\right)  }%
\exp\left(  \frac{-c\left(  m_{j}^{2}c^{2}+p_{_{\perp}0(j)}^{2}+p_{_{\parallel
(j)}}^{2}\right)  ^{\frac{1}{2}}}{T_{z(j)}}\right)  \tag{7}%
\end{equation}
Let us define follwoing identity%
\begin{equation}
\widehat{\gamma}_{\perp0(j)}=\left[  1+\frac{p_{\perp0(j)}^{2}}{m_{j}^{2}%
c^{2}}\right]  ^{1/2}\tag{8}%
\end{equation}
and so
\begin{equation}
\gamma_{\perp0(j)}=\widehat{\gamma}_{\perp0(j)}\left[  1+\frac{p_{\parallel
(j)}^{2}}{\left(  m_{j}^{2}c^{2}\right)  \widehat{\gamma}_{\perp0}^{2}%
}\right]  ^{1/2}\tag{9}%
\end{equation}
introducing the new variables such as
\begin{equation}
p_{_{\parallel(j)}}=\left(  \widehat{\gamma}_{\perp0(j)}m_{j}c\right)
\sinh\theta\tag{10}%
\end{equation}
Eqs. (8) - (9) give us
\begin{equation}
\gamma_{0(j)}=\left(  \widehat{\gamma}_{\perp0(j)}\right)  \cosh\theta;\text{
}\frac{dp_{_{\parallel(j)}}}{\gamma_{0(j)}}=\left(  m_{j}c\right)
d\theta\tag{11}%
\end{equation}
and by using the identity
\begin{equation}
\int_{0}^{\infty}\tau d\tau\exp\left[  i\left(  \gamma_{0(j)}\omega
-\frac{k_{z}p_{\perp0(j)}}{m_{j}}\right)  \tau\right]  =-\frac{1}{\left(
\gamma_{0(j)}\omega-\frac{k_{z}p_{\perp0(j)}}{m_{j}}\right)  ^{2}}\tag{12}%
\end{equation}
substitution of Eqs. (7) and $(10)-(12)$ into (6) and integrating over
$\theta$ gives us the dispersion relation
\begin{equation}
0=1-\frac{c^{2}k_{z}^{2}}{\omega^{2}}-\sum_{j}\frac{\omega_{p_{j}}^{2}}%
{\omega^{2}\widehat{\gamma}_{\perp_{j}}}\left[  \frac{K_{0}\left(  \frac
{m_{j}c^{2}}{T_{z(j)}}\widehat{\gamma}_{\perp}\right)  }{K_{1}\left(
\frac{m_{j}c^{2}}{T_{z(j)}}\right)  }+\frac{p_{\perp0(j)}^{2}\left(
\omega^{2}-c^{2}k^{2}\right)  }{K_{1}\left(  \frac{m_{j}c^{2}}{T_{z(j)}%
}\right)  2m_{j}^{2}c^{2}\widehat{\gamma}_{\perp(j)}^{2}}\int_{0}^{\infty
}K_{0}(\eta)\tau d\tau\right]  \tag{13}%
\end{equation}
where argument $\eta=\left[  \left(  \frac{m_{j}c^{2}}{T_{z(j)}}%
\widehat{\gamma}_{\perp0(j)}-i\omega\tau\right)  ^{2}+\left(  ck_{z}%
\tau\right)  ^{2}\right]  ^{1/2}.$ For deriving (13), we have used the
identity []%
\begin{equation}
\frac{1}{\pi}K_{0}[\left(  a^{2}+b^{2}\right)  ^{1/2}]=\frac{1}{2\pi}%
\int_{-\infty}^{+\infty}d\theta\exp(-ib\sinh\theta-a\cosh\theta)\tag{14}%
\end{equation}
condition \qquad\ , allows us do Taylor expansion of $K_{0}$
\begin{align}%
{\displaystyle\int\limits_{0}^{\infty}}
\tau d\tau K_{0}(\eta) &  =%
{\displaystyle\int\limits_{0}^{\infty}}
\tau d\tau K_{0}(\beta_{\perp(j)}^{2}+\sigma^{2}\tau^{2})^{\frac{1}{2}%
}\nonumber\\
&  +%
{\displaystyle\int\limits_{0}^{\infty}}
\tau d\tau\frac{i\beta_{\perp(j)}\omega\tau}{(\beta_{\perp(j)}^{2}+\sigma
^{2}\tau^{2})^{\frac{1}{2}}}K_{1}(\beta_{\perp(j)}^{2}+\sigma^{2}\tau
^{2})^{\frac{1}{2}})\tag{15}%
\end{align}
where we have denoted $\frac{m_{j}c^{2}}{T_{z(j)}}\widehat{\gamma}_{\perp
}=\beta_{\perp(j)};$ $\left(  c^{2}k_{z}^{2}-\omega^{2}\right)  ^{1/2}=\sigma$
and have used the identity $dK_{0}/dx=-K_{1}(x).$ Substituting (15) into (13)
and performing the integrations gives us follwoing general disperison relation
for the propagation of electromagnetic waves in electron positron ion plasma
where electrons and positrons are active whereas ions form a neutralizing
background.
\begin{equation}
0=1-\frac{c^{2}k_{z}^{2}}{\omega^{2}}-\sum_{j}\frac{\omega_{pj}^{2}}%
{\omega^{2}\widehat{\gamma}_{\perp(j)}}\left[
\begin{array}
[c]{c}%
\frac{K_{0}(\beta_{\perp(j)})}{K_{1}(\beta_{(j)})}-\frac{p_{\perp0(j)}%
^{2}K_{1}(\beta_{\perp(j)})}{2m_{j}T_{z(j)}\widehat{\gamma}_{\perp(j)}%
K_{1}(\beta_{(j)})}\\
-\frac{p_{\perp0(j)}^{2}(\beta_{\perp(j)})^{\frac{1}{2}}K_{\frac{1}{2}}%
(\beta_{\perp(j)})}{2m_{j}\widehat{\gamma}_{\perp(j)}T_{z(j)}K_{1}(\beta
_{(j)})}\frac{i\omega\left(  \frac{\pi}{2}\right)  ^{\frac{1}{2}}}{(c^{2}%
k_{z}^{2}-\omega^{2})^{\frac{1}{2}}}%
\end{array}
\right]  \tag{16}%
\end{equation}
To obatin above relation, we have used the follwoing MacDonald function
integrals [ ]
\begin{equation}%
{\displaystyle\int\limits_{0}^{\infty}}
\tau d\tau K_{0}(\beta_{\perp(j)}^{2}+\sigma^{2}\tau^{2})^{\frac{1}{2}}%
=\frac{\beta_{\perp(j)}}{(c^{2}k_{z}^{2}-\omega^{2})}K_{1}(\beta_{\perp
(j)})\tag{17}%
\end{equation}
and%
\begin{align}
&
{\displaystyle\int\limits_{0}^{\infty}}
\tau d\tau\frac{i\beta_{\perp(j)}\omega\tau}{(\beta_{\perp(j)}^{2}+\sigma
^{2}\tau^{2})^{\frac{1}{2}}}K_{1}(\beta_{\perp(j)}^{2}+\sigma^{2}\tau
^{2})^{\frac{1}{2}})\nonumber\\
&  =\left[  \frac{i\omega(\beta_{\perp(j)})^{\frac{3}{2}}(\frac{\pi}%
{2})^{\frac{1}{2}}}{(c^{2}k_{z}^{2}-\omega^{2})(c^{2}k_{z}^{2}-\omega
^{2})^{\frac{1}{2}}}\right]  K_{\frac{1}{2}}(\beta_{\perp(j)})\tag{18}%
\end{align}

\section{ Analysis of electromagnetic instability}

The dispersion relation (16) is complex or purely imaginary, there is a
possibility of obtaining unstable situation. In the subsequent section we
investigate various situations for the interaction of relativistic EM pump
wave with electron and positron particles. ELECTRON POSITRON

\paragraph{\textbf{Case 1: }EMW wave with large relativistic transverse
energy}

We first investigate a situation in which the relativistic transverse energy
of the electromagnetic pump wave is larger than its thermal renergy
$\frac{m_{j}c^{2}}{T_{z(j)}}\widehat{\gamma}_{\perp}\gg1$ or $\left(
\beta_{\perp(j)}\gg1\right)  .$ This allows us to expand the MacDonald
function $K_{0}$, $K_{1}$, and $K_{1/2}$ in the follwoing manner%
\begin{equation}
K_{0}\left(  \frac{m_{j}c^{2}}{T_{z(j)}}\widehat{\gamma}_{\perp}\right)
\approx\left(  \frac{\pi}{2\beta_{\perp(j)}}\right)  ^{1/2}e^{\left(
\frac{m_{j}c^{2}}{T_{z(j)}}\widehat{\gamma}_{\perp}\right)  }\approx
K_{1}\left(  \frac{m_{j}c^{2}}{T_{z(j)}}\widehat{\gamma}_{\perp}\right)
\approx K_{\frac{1}{2}}\left(  \frac{m_{j}c^{2}}{T_{z(j)}}\widehat{\gamma
}_{\perp}\right)  \tag{19}%
\end{equation}
Neglecting the higher order terms we have%
\begin{equation}
\frac{K_{\frac{1}{2}}\left(  \frac{m_{j}c^{2}}{T_{z(j)}}\widehat{\gamma
}_{\perp}\right)  }{K_{1}\left(  \frac{m_{j}c^{2}}{T_{z(j)}}\widehat{\gamma
}_{\perp}\right)  }\approx\frac{K_{0}\left(  \frac{m_{j}c^{2}}{T_{z(j)}%
}\widehat{\gamma}_{\perp}\right)  }{K_{1}\left(  \frac{m_{j}c^{2}}{T_{z(j)}%
}\widehat{\gamma}_{\perp}\right)  }\approx\frac{1}{\widehat{\gamma}_{\perp
}^{\frac{1}{2}}}e^{\frac{m_{j}c^{2}(1-\widehat{\gamma}_{\perp})}{T_{z(j)}}}
\tag{20}%
\end{equation}
Inserting Eq. 21 in Eq. 20 we obtain%
\begin{equation}
0=1-\frac{c^{2}k_{z}^{2}}{\omega^{2}}-\sum_{j}\frac{\omega_{pj}^{2}}%
{\omega^{2}\widehat{\gamma}_{\perp}}\left[
\begin{array}
[c]{c}%
\frac{e^{\frac{m_{j}c^{2}(1-\widehat{\gamma}_{\perp})}{T_{z(j)}}}}%
{\widehat{\gamma}_{\perp}^{\frac{1}{2}}}\left(  1-\frac{m_{j}c^{2}%
(\widehat{\gamma}_{\perp}^{2}-1)}{2T_{z(j)}\widehat{\gamma}_{\perp}}\right) \\
-i\omega\left(  \frac{m_{j}c^{2}}{2T_{z(j)}}\right)  ^{3/2}\left(  \frac{\pi
}{c^{2}k_{z}^{2}-\omega^{2}}\right)  ^{1/2}\frac{(\widehat{\gamma}_{\perp}%
^{2}-1)e^{\frac{m_{j}c^{2}(1-\widehat{\gamma}_{\perp})}{T_{z(j)}}}}%
{\widehat{\gamma}_{\perp}}%
\end{array}
\right]  \tag{21}%
\end{equation}
here we have defined effective momentum of the pumping wave
\[
m_{j}^{2}c^{2}(\widehat{\gamma}_{\perp}^{2}-1)=p_{\perp0(j)}^{2}%
\]
Below we discuss above equation for the the low and high frequency cases, respectively.

\paragraph{\textbf{Low frequency regime}}

In the present section, we solve Eq. 21 for low frequency case, i.e., when
$(c^{2}k_{z}^{2}-\omega^{2})^{1/2}\thickapprox ck_{z}$ Under these conditions,
Eq. 22 can be simplified and we obtain for a purely imaginary expression given
by%
\begin{equation}
\operatorname{Im}\omega=-\frac{8k_{z}\left(  \frac{T_{+}}{m_{+}}\right)
^{3/2}}{\pi^{1/2}c^{2}(\widehat{\gamma}_{\perp}^{2}-1)}\frac{\left[
\begin{array}
[c]{c}%
\frac{c^{2}k_{z}^{2}\widehat{\gamma}_{\perp}^{3/2}}{\omega_{p+}^{2}%
e^{\beta_{+}(1-\widehat{\gamma}_{\perp})}}+\left(  1+\left(  \frac{n_{-}%
}{n_{+}}\frac{m_{+}}{m-}\right)  e^{\left(  \beta_{+}-\beta_{-}\right)
(1-\widehat{\gamma}_{\perp})}\right)  \\
-\frac{\beta_{+}(\widehat{\gamma}_{\perp}^{2}-1)}{2\widehat{\gamma}_{\perp}%
}\left(  1+\left(  \frac{n_{-}}{n_{+}}\frac{m_{+}}{m-}\right)  \left(
\frac{T_{+}}{T-}\right)  ^{3/2}e^{\left(  \beta_{+}-\beta_{-}\right)
(1-\widehat{\gamma}_{\perp})}\right)
\end{array}
\right]  }{\left[  1+\left(  \frac{n_{-}}{n_{+}}\frac{m_{+}}{m-}\right)
\left(  \frac{T_{+}}{T-}\right)  ^{3/2}e^{\left(  \beta_{+}-\beta_{-}\right)
(1-\widehat{\gamma}_{\perp})}\right]  }\tag{22}%
\end{equation}
let $\omega_{pp}^{2\ast}=\omega_{pp}^{2}\frac{\exp\beta_{+}(1-\widehat{\gamma
}_{\perp})}{\widehat{\gamma}_{\perp}^{3/2}}$%
\begin{equation}
\operatorname{Im}\omega=-\frac{k_{z}\left(  \frac{2T_{+}}{m_{+}}\right)
^{3/2}}{\pi^{1/2}c^{2}(\widehat{\gamma}_{\perp}^{2}-1)}\frac{\left[
\begin{array}
[c]{c}%
\frac{c^{2}k_{z}^{2}}{\omega_{p+}^{2\ast}}+\left(  1+\left(  \frac{n_{-}%
}{n_{+}}\frac{m_{+}}{m-}\right)  e^{\left(  \beta_{+}-\beta_{-}\right)
(1-\widehat{\gamma}_{\perp})}\right)  \\
-\frac{\beta_{+}(\widehat{\gamma}_{\perp}^{2}-1)}{2\widehat{\gamma}_{\perp}%
}\left(  1+\left(  \frac{n_{-}}{n_{+}}\frac{m_{+}}{m-}\right)  \left(
\frac{T_{+}}{T-}\right)  ^{3/2}e^{\left(  \beta_{+}-\beta_{-}\right)
(1-\widehat{\gamma}_{\perp})}\right)
\end{array}
\right]  }{\left[  1+\left(  \frac{n_{-}}{n_{+}}\frac{m_{+}}{m-}\right)
\left(  \frac{T_{+}}{T-}\right)  ^{3/2}e^{\left(  \beta_{+}-\beta_{-}\right)
(1-\widehat{\gamma}_{\perp})}\right]  }\tag{23}%
\end{equation}%
\begin{equation}
\operatorname{Im}\omega=-\frac{k_{z}\left(  \frac{2T_{+}}{m_{+}}\right)
^{3/2}}{\pi^{1/2}c^{2}(\widehat{\gamma}_{\perp}^{2}-1)}\frac{\left[
\frac{c^{2}k_{z}^{2}}{\omega_{p+}^{2\ast}}+\left(
\begin{array}
[c]{c}%
1-\frac{\beta_{+}(\widehat{\gamma}_{\perp}^{2}-1)}{2\widehat{\gamma}_{\perp}%
}\\
+\left(  \frac{n_{-}}{n_{+}}\frac{m_{+}}{m-}\right)  e^{\left(  \beta
_{+}-\beta_{-}\right)  (1-\widehat{\gamma}_{\perp})}(1-\frac{\beta
_{+}(\widehat{\gamma}_{\perp}^{2}-1)}{2\widehat{\gamma}_{\perp}}\left(
\frac{T_{+}}{T-}\right)  ^{3/2})
\end{array}
\right)  \right]  }{\left[  1+\left(  \frac{n_{-}}{n_{+}}\frac{m_{+}}%
{m-}\right)  \left(  \frac{T_{+}}{T-}\right)  ^{3/2}e^{\left(  \beta_{+}%
-\beta_{-}\right)  (1-\widehat{\gamma}_{\perp})}\right]  }\tag{24}%
\end{equation}
\textbf{High frequency regime}

In this section we investigate the high frequency regime, when $\omega\gg
ck_{z}$ which implies that $(c^{2}k_{z}^{2}-\omega^{2})^{\frac{1}{2}}=i\omega$
. After simplifying Eq. 22 , we get
\begin{equation}
\operatorname{Im}\omega=\left(  \frac{\omega_{pp}}{\widehat{\gamma}_{\perp}%
}e^{\frac{\beta_{p}(1-\widehat{\gamma}_{\perp})}{2}}\right)  \left[
\begin{array}
[c]{c}%
\frac{(2\widehat{\gamma}_{\perp}-\beta_{p}(\widehat{\gamma}_{\perp}^{2}%
-1))}{2\widehat{\gamma}_{\perp}^{1/2}}\left\{  1+(\frac{n_{e}}{n_{p}}%
)(\frac{m_{p}}{m_{e}})\left(  \frac{(2\widehat{\gamma}_{\perp}-\beta
_{e}(\widehat{\gamma}_{\perp}^{2}-1))}{(2\widehat{\gamma}_{\perp}-\beta
_{p}(\widehat{\gamma}_{\perp}^{2}-1))}\right)  e^{(\beta_{e}-\beta
_{p})(1-\widehat{\gamma}_{\perp})}\right\}  \\
-\left(  \frac{\pi}{2}\right)  ^{1/2}\beta_{p}^{3/2}(\widehat{\gamma}_{\perp
}^{2}-1)\left\{  1+(\frac{n_{e}}{n_{p}})(\frac{m_{p}}{m_{e}})^{5/2}\left(
\frac{T_{p}}{T_{e}}\right)  ^{3/2}e^{(\beta_{e}-\beta_{p})(1-\widehat{\gamma
}_{\perp})}\right\}
\end{array}
\right]  ^{1/2}\tag{25}%
\end{equation}
or%
\begin{equation}
\operatorname{Im}\omega=\left(  \frac{\omega_{pp}}{\widehat{\gamma}_{\perp}%
}e^{\frac{\beta_{p}(1-\widehat{\gamma}_{\perp})}{2}}\right)  \left[
\begin{array}
[c]{c}%
\frac{(2\widehat{\gamma}_{\perp}-\beta_{p}(\widehat{\gamma}_{\perp}^{2}%
-1))}{2\widehat{\gamma}_{\perp}^{1/2}}\left\{  1+(\frac{n_{e}}{n_{p}}%
)(\frac{m_{p}}{m_{e}})\left(  \frac{(2\widehat{\gamma}_{\perp}-\beta
_{e}(\widehat{\gamma}_{\perp}^{2}-1))}{(2\widehat{\gamma}_{\perp}-\beta
_{p}(\widehat{\gamma}_{\perp}^{2}-1))}\right)  e^{(\beta_{e}-\beta
_{p})(1-\widehat{\gamma}_{\perp})}\right\}  \\
-\left(  \frac{\pi}{2}\right)  ^{1/2}\beta_{p}^{3/2}(\widehat{\gamma}_{\perp
}^{2}-1)\left\{  1+(\frac{n_{e}}{n_{p}})(\frac{m_{p}}{m_{e}})^{5/2}\left(
\frac{T_{p}}{T_{e}}\right)  ^{3/2}e^{(\beta_{e}-\beta_{p})(1-\widehat{\gamma
}_{\perp})}\right\}
\end{array}
\right]  ^{1/2}\tag{26}%
\end{equation}
let%
\[
\left(  \frac{(2\widehat{\gamma}_{\perp}-\beta_{e}(\widehat{\gamma}_{\perp
}^{2}-1))}{(2\widehat{\gamma}_{\perp}-\beta_{p}(\widehat{\gamma}_{\perp}%
^{2}-1))}\right)  =\frac{a}{b}%
\]%
\[
e^{(\beta_{e}-\beta_{p})(1-\widehat{\gamma}_{\perp})}=d
\]%
\begin{equation}
\operatorname{Im}\omega=\left(  \frac{\omega_{pp}}{\widehat{\gamma}_{\perp}%
}e^{\frac{\beta_{p}(1-\widehat{\gamma}_{\perp})}{2}}\right)  \left[
\begin{array}
[c]{c}%
\frac{b}{2\widehat{\gamma}_{\perp}^{1/2}}\left\{  1+(\frac{n_{e}}{n_{p}%
})(\frac{m_{p}}{m_{e}})(\frac{da}{b})\right\}  \\
-\left(  \frac{\pi}{2}\right)  ^{1/2}\beta_{p}^{3/2}(\widehat{\gamma}_{\perp
}^{2}-1)\left\{  1+d(\frac{n_{e}}{n_{p}})(\frac{m_{p}}{m_{e}})^{5/2}\left(
\frac{T_{p}}{T_{e}}\right)  ^{3/2}\right\}
\end{array}
\right]  ^{1/2}\tag{27}%
\end{equation}
In Fig. 2 we have considered the positive root of Eq. (25). \ \ 

\subsubsection{\textbf{Case: Large thermal relativistic energy }}

In present section we consider the opposite limit to that investigated in Sec.
III. That is the case when the thermal relativistic energy is large as
compared to the transverse relativistic energy of the pump wave. Thus, here we
have%
\[
\beta_{\perp}\ll1,\beta\ll1
\]
and subsequently, we have the following relationships for the MacDonald
functions:%
\begin{equation}
K_{1}(\beta_{1})\approx\frac{1}{\beta_{\perp}},\text{ }K_{1}(\beta
)\approx\frac{1}{\beta},\text{ }K_{0}(\beta_{\perp})\approx-\ln\frac
{\beta_{\perp}}{2},K_{\frac{1}{2}}(\beta_{\perp})\approx\sqrt{\frac{\pi
}{2\beta_{\perp}}}\tag{28}%
\end{equation}
Inserting Eq. 26 in the general dispersion relation Eq. 20, we get%
\begin{align}
0 &  =1-\frac{c^{2}k_{z}^{2}}{\omega^{2}}-\frac{\omega_{pp}^{2}}{\omega
^{2}\widehat{\gamma}_{\perp}}\left[  -\beta^{p}\ln(\frac{\beta_{\perp}^{p}}%
{2})-\frac{\beta^{p}(\widehat{\gamma}_{\perp}^{2}-1)}{2\widehat{\gamma}%
_{\perp}}-\frac{\beta^{p}(\widehat{\gamma}_{\perp}^{2}-1)i\omega\left(
\frac{\pi}{2}\right)  ^{\frac{1}{2}}}{2\widehat{\gamma}_{\perp}(c^{2}k_{z}%
^{2}-\omega^{2})^{\frac{1}{2}}}\right]  \nonumber\\
&  -\frac{\omega_{pe}^{2}}{\omega^{2}\widehat{\gamma}_{\perp}}\left[
-\beta^{e}\ln(\frac{\beta_{\perp}^{e}}{2})-\frac{\beta^{e}(\widehat{\gamma
}_{\perp}^{2}-1)}{2\widehat{\gamma}_{\perp}}-\frac{\beta^{e}(\widehat{\gamma
}_{\perp}^{2}-1)i\omega\left(  \frac{\pi}{2}\right)  ^{\frac{1}{2}}}%
{2\widehat{\gamma}_{\perp}(c^{2}k_{z}^{2}-\omega^{2})^{\frac{1}{2}}}\right]
\tag{29}%
\end{align}
As previously, we separately investigate the low and high frequency regimes.

\subsubsection{\textbf{Low frequency regime}}

Now we investigate Eq. (27) for the low frequency case which reduces to an
expression for purely imaginary . %

\begin{equation}
\operatorname{Im}\omega=-\frac{\left[  c^{2}k_{z}^{2}-\frac{\omega_{pp}^{2}%
}{\widehat{\gamma}_{\perp}}\beta^{p}\ln\left(  \frac{\beta_{\perp}^{p}}%
{2}\right)  \left[  1+(\frac{n_{e}}{n_{p}})(\frac{m_{p}}{m_{e}})\ln\left(
\frac{\beta_{\perp}^{e}}{\beta_{\perp}^{p}}\right)  \right]  -\frac
{\omega_{pp}^{2}\beta_{\perp}^{p}(\widehat{\gamma}_{\perp}^{2}-1)}%
{2\widehat{\gamma}_{\perp}^{3}}\left(  1+(\frac{T_{p}}{T_{e}})(\frac{n_{e}%
}{n_{p}})^{3/2}(\frac{m_{p}}{m_{e}})^{5/2}\right)  \right]  }{\left[
\frac{\pi\omega_{pp}^{2}(\widehat{\gamma}_{\perp}^{2}-1)\beta^{p^{2}}%
}{4\widehat{\gamma}_{\perp}^{2}}\left(  1+(\frac{T_{p}}{T_{e}})^{2}%
(\frac{m_{e}}{m_{p}})^{2}\right)  \right]  } \tag{30}%
\end{equation}

\subsubsection{\textbf{High frequency regime}}%

\[
\omega=\pm\frac{\omega_{pp}}{\widehat{\gamma}_{\perp}^{1/2}}\left[  -\beta
^{p}\ln\left(  \frac{\beta_{\perp}^{p}}{2}\right)  \left(  1+(\frac{n_{e}%
}{n_{p}})(\frac{T_{p}}{T_{e}})\right)  -\frac{\beta^{p}(\widehat{\gamma
}_{\perp}^{2}-1)}{2\widehat{\gamma}_{\perp}}\left(  1+(\frac{n_{e}}{n_{p}%
})(\frac{T_{p}}{T_{e}})\right)  +(\frac{n_{e}}{n_{p}})(\frac{T_{p}}{T_{e}%
})\beta^{e}+\frac{\pi\widehat{\gamma}_{\perp}\beta^{p}}{2}\right]  ^{1/2}
\]
$\omega_{r}=\operatorname{Re}(\omega)$%
\[
\operatorname{Re}(\omega)=\frac{\omega_{pp}}{\widehat{\gamma}_{\perp}^{1/2}%
}\left[  -\beta^{p}\ln\left(  \frac{\beta_{\perp}^{p}}{2}\right)  \left(
1+(\frac{n_{e}}{n_{p}})(\frac{T_{p}}{T_{e}})\right)  -\frac{\beta^{p}%
(\widehat{\gamma}_{\perp}^{2}-1)}{2\widehat{\gamma}_{\perp}}\left(
1+(\frac{n_{e}}{n_{p}})(\frac{T_{p}}{T_{e}})\right)  +(\frac{n_{e}}{n_{p}%
})(\frac{T_{p}}{T_{e}})\beta^{e}+\frac{\pi\widehat{\gamma}_{\perp}\beta^{p}%
}{2}\right]  ^{1/2}
\]
and
\[
\operatorname{Im}\omega=\frac{\omega_{pp}}{\widehat{\gamma}_{\perp}^{1/2}%
}\left[  \beta^{p}\ln\left(  \frac{\beta_{\perp}^{p}}{2}\right)  \left(
1+(\frac{n_{e}}{n_{p}})(\frac{T_{p}}{T_{e}})\right)  +\frac{\beta^{p}%
(\widehat{\gamma}_{\perp}^{2}-1)}{2\widehat{\gamma}_{\perp}}\left(
1+(\frac{n_{e}}{n_{p}})(\frac{T_{p}}{T_{e}})\right)  +(\frac{n_{e}}{n_{p}%
})(\frac{T_{p}}{T_{e}})\beta^{e}+\frac{\pi\widehat{\gamma}_{\perp}\beta^{p}%
}{2}\right]  ^{1/2}
\]

\section{Conclusions }

In this paper we have derived a general dispersion relation for
electromagnetic waves in a single component relativistic plasma where the
transverse energy of the wave is relativistic and also thermal energy is
relativistic. The dynamics of the ions has been neglected. This general
dispersion relation has been investigated in detail for three limiting cases.
First when the transverse energy of the electromagnetic wave is large in
comparison with the thermal energy, which in the first instance has been taken
as being non-relativistic; second when the thermal energy is large and
relativistic in comparison with the wave transverse energy; and third for
large relativistic transverse and relativistic thermal energies. Each of these
cases is investigated separately for the high and low frequency regimes.
Expressions are obtained for the growth/damping rates of the possible
instability and the dependence of the growth rates on the different defining
parameters has also been investigated and plots obtained for these different
cases.

\end{document}